\documentclass[runningheads]{llncs}

\let\vec\relax
\DeclareMathAccent{\vec}{\mathord}{letters}{"7E}
\usepackage[utf8]{inputenc}
\usepackage{pgfplots}
\DeclareUnicodeCharacter{2212}{−}
\usepgfplotslibrary{groupplots,dateplot}
\usetikzlibrary{patterns,shapes.arrows}
\pgfplotsset{compat=newest}

\usepackage{xspace}
\usepackage{mathtools}
\usepackage{nicefrac}
\usepackage{amsfonts}
\usepackage{pifont}
\usepackage{floatflt}
\usepackage{url}
\usepackage{dsfont}

\usepackage{tikz}
\usetikzlibrary{calc, positioning, shadows, backgrounds, arrows.meta}
\usepackage{multirow}
\usepackage{booktabs}
\usepackage{tabularx}
\newcolumntype{L}{>{\raggedright\arraybackslash}X}

\usepackage{algorithm}

\makeatletter
\DeclareRobustCommand*\cal{\@fontswitch\relax\mathcal}
\makeatother

\newcommand{\ie}{\emph{i.e.,}\xspace} 
\newcommand{\eg}{\emph{e.g.,}\xspace} 

\newcommand{\mypara}[1]{%
  \noindent\textbf{#1. }\ignorespaces
}

\newcommand{\commentout}[1]{}

\newcommand{\V}{{\cal V}}

\newcommand{\sat}{\models}

\newcommand{\lem}{\begin{lemma}}
\newcommand{\elem}{\end{lemma}}
\newcommand{\pro}{\begin{proposition}}
\newcommand{\epro}{\end{proposition}}

\newcommand{\dfn}{\begin{definition}}
\newcommand{\edfn}{\end{definition}}

\newcommand{\xam}{\begin{example}}
\newcommand{\exam}{\end{example}}

\usepackage{algorithm}

\newcommand{\SoAno}{8\xspace}

\usepackage{longtable}
\usepackage{pgfplots}
\usepackage{pgfplotstable}
\usepackage{paralist}
\usepgfplotslibrary{fillbetween}
\pgfplotsset{compat=1.16}

\usepackage[T1]{fontenc}
\usepackage{listings}
\usepackage{graphicx}
\usepackage{xspace}
\usepackage{booktabs}
\usepackage[hidelinks]{hyperref}
\usepackage{cleveref}
\usepackage{color}
\usepackage{todonotes}
\usepackage{algorithm}
\usepackage[noend]{algpseudocode}
\algrenewcommand\algorithmicindent{0.8em}
\usepackage[shortlabels]{enumitem}
\usepackage{tikz}
\usepackage{pgfplots}
\usepackage{subcaption}

\pgfplotsset{compat=1.18,
        /pgfplots/ybar legend/.style={
        /pgfplots/legend image code/.code={%
        \draw[##1,/tikz/.cd,bar width=3pt,yshift=-0.2em,bar shift=0pt]
                plot coordinates {(0cm,0.8em)};},
},
}
\definecolor{colBase}{HTML}{73726C}
\definecolor{colDistH}{HTML}{2A78D6}
\definecolor{colDistS}{HTML}{1BAF7A}
\definecolor{colPen}{HTML}{EB6834}
\definecolor{colPenDistH}{HTML}{EDA100}
\definecolor{colPenDistS}{HTML}{C026D3}

\lstdefinestyle{mystyle}{
    backgroundcolor=\color{backcolour},   
    commentstyle=\color{codegreen},
    keywordstyle=\color{magenta},
    numberstyle=\tiny\color{codegray},
    stringstyle=\color{codepurple},
    basicstyle=\ttfamily\footnotesize,
    breakatwhitespace=false,         
    breaklines=true,                 
    captionpos=b,                    
    keepspaces=true,                 
    numbers=left,                    
    numbersep=5pt,                  
    showspaces=false,                
    showstringspaces=false,
    showtabs=false,                  
    tabsize=2
}

\newcommand{\probP}{\text{I\kern-0.15em P}}
\newcommand{\sectref}[1]{%
  \hyperref[#1]{Section~\ref{#1}}%
}
\newcommand{\subsectref}[1]{%
  \hyperref[#1]{Subsection~\ref{#1}}%
}
\crefformat{section}{§#2#1#3}
\Crefformat{section}{§#2#1#3}

\newcommand{\tool}{\textsc{s}oft\textsc{r}e\textsc{x}\xspace}
\newcommand{\softrex}[1]{\textsc{s}oft\textsc{r}e\textsc{x}\xspace} 

\newcommand{\rexbase}[1]{$\softrex\_{Base}$\xspace}
\newcommand{\rexpenalty}[1]{\textit{\softrex\_{Penalty}}\xspace}
\newcommand{\rexdist}[1]{\textit{\softrex\_{Dist}}\xspace}
\newcommand{\rexdistsem}[1]{\textit{\softrex\_{DistSem}}\xspace}
\newcommand{\rexpendistsem}[1]{\textit{\softrex\_{Penalty+DistSem}}\xspace}
\newcommand{\rexpendist}[1]{\textit{\softrex\_{Penalty+Dist}}\xspace}

\newcommand{\muse}[1]{\textsc{Muse}\xspace}
\newcommand{\metallaxis}[1]{\textsc{Metallaxis}\xspace}

\begin{document}

\title{Causality-Based Fault Localization}

%
%
\author{Melane Navaratnarajah\inst{1}\orcidID{0009-0001-8987-6134} \and
Aidan Dakhama\inst{2}\orcidID{0009-0002-7318-7964} \and
Tom Yaacov\inst{1}\orcidID{0000-0002-0565-6506}\and 
David A. Kelly\inst{1}\orcidID{0000-0002-5368-6769} \and 
Karine Even-Mendoza \inst{1}\orcidID{0000-0002-3099-1189} \and 
Hana Chockler\inst{1}\orcidID{0000-0003-1219-0713}}
%
\authorrunning{M. Navaratnarajah et al.}
%
\institute{King’s College London, London, UK\\
\email{\{melane.navaratnarajah,tom.yaacov,david.a.kelly,\\ karine.even\_mendoza,hana.chockler\}@kcl.ac.uk}\and
University of Edinburgh, Edinburgh, UK\\
\email{aidandakhama@ed.ac.uk}}
%
\maketitle              

\begin{abstract}
    We introduce \emph{Causality-Based Fault Localization (CBFL)}, a novel approach to fault localization grounded in the theory of \textit{actual causality}. We use the notion of \emph{causal responsibility} to formalize and define the contribution of a given location in the source code under the context of a failing test. We also utilize a distance metric to prioritize minimal fixes and penalize broken tests, leading to more interpretable and fixable fault localization. We implemented CBFL in a tool, \tool, and our evaluation results demonstrate that \tool outperforms state-of-the-art \textit{Mutation-Based Fault Localization} (MBFL) tools based on \textit{Mean Reciprocal Rank} (MRR) and Top-1 by $16\%$ on both metrics, when using the same set of mutants. We also propose a sophisticated mutation generation strategy based on the structure and representation of the program to improve the accuracy of our ranked list, which outperforms state-of-the-art methods by $2.6\times$ in terms of required mutants evaluated while remaining competitive in performance.

\end{abstract}

\keywords{Causality, Software Engineering, Fault Localization, Testing}

\section{Introduction}


\mypara{Motivation} Software debugging remains one of the most challenging and costly activities in software engineering~\cite{Wong2016Survey}. As software
systems continue to grow in scale and complexity, and with the increase in LLM-generated commits, 
the hard problem of identifying the root causes of failures is only becoming increasingly difficult. Finding the error is often harder than fixing: establishing where errors originate often dominates the effort required to correct the problem~\cite{Kang2024AutoFL}.
\textit{Fault Localization} (FL), the task of identifying the program elements responsible for a failure, is widely recognized as one of the most time-consuming and
expensive stages of debugging and has consequently been the subject of extensive research for more than two decades.
Despite decades of research and substantial progress in automated FL, accurately identifying the source of an observed failure remains a major challenge. Recent studies continue to highlight the effort associated with FL and the limitations of existing techniques, motivating ongoing work on more effective debugging support~\cite{Wong2016Survey,Kang2024AutoFL}.

\mypara{State of the art} Among automated FL techniques, \textit{Spectrum-Based Fault Localization} (SBFL) has emerged as one of the most widely studied and influential paradigms~\cite{abreu2007accuracy}. 
SBFL analyzes program spectra---information about which program elements are executed during a particular run, such as statement or branch coverage collected from passing and failing executions---and assigns each element a suspiciousness score based on its statistical association with observed failures~\cite{jones2005empirical,abreu2009practical}.
Numerous ranking metrics have been proposed, including Tarantula~\cite{jones2005empirical},
Muse~\cite{moon_ask_2014}, and DStar~\cite{6651713}, with empirical studies demonstrating that SBFL substantially reduces the search space developers
must inspect during debugging. More recently,
\textit{Mutation-Based Fault Localization} (MBFL) 
has extended this approach by replacing observation with \emph{active intervention}, generating mutants and analyzing their impact on test outcomes to infer fault relevance. Techniques such as Metallaxis~\cite{papadakisMetallaxisFLMutationbasedFault2015b} have shown that mutation information can improve localization accuracy over traditional spectrum-based approaches, while frameworks such as Major~\cite{major} enabled efficient large-scale mutation analysis.

\mypara{Limitations of SBFL and MBFL} Despite their practical effectiveness, both MBFL and SBFL remain
fundamentally associational: program elements are ranked according to correlations with failures across executions rather than their causal contribution to the observed failure. But the question being asked is causal: 
\textit{what and where is the cause of the failure?}
As a result of its associational nature, highly ranked program elements that are not necessarily causes of the failure remain under investigation. In addition to this wasted effort, current methods struggle to distinguish between cases that only causal methods can address: over-determination (multiple faults each sufficient to produce a failure), pre-emption (one fault causing an observed failure before another fault occurs), and error masking (the effects of a fault being hidden or corrected before reaching the output). 
This raises a fundamental question: \emph{to what extent can FL benefit from causal reasoning, rather than statistical association alone?}

\mypara{Contribution} In this paper, we formalize FL as a causal analysis problem. We use the theory of actual causality~\cite{halpern_actual_2016} to construct \textit{Structural Causal Models} (SCMs) reflecting
the underlying structure of the program. In our formulation, the states of individual code elements act as variables, and mutations serve as targeted interventions. By observing the test suite executions of a program under these interventions, we are able to quantify the causal contribution or degree of responsibility~\cite{chockler2004responsibility} of any code element toward a test failure. Crucially, our SCM allows us to navigate masked and overdetermined faults by evaluating interventions, while still admitting efficient computation and more accuracy.

To make this theoretical framework computationally tractable on real-world software, we introduce \tool{}, a causality-based FL technique. Rather than relying on exhaustive, brute force mutation generation, \tool{} allows for some sampling of mutants or employs a hierarchical search over the \textit{Concrete Syntax Tree} (CST) to dynamically iterate over potential minimal sets of code elements to mutate--contingency sets. To ensure practical accuracy, \tool{} regularizes its base causal responsibility scores by penalizing interventions that break originally passing tests, \textit{passing-to-failing} (P2F) penalization, and weighing them against the syntactic or semantic distance of the mutation.

We evaluate \tool{} on the Defects4J benchmark \cite{just_defects4j_2014}, a standard benchmark for reproducible bugs, against 8 state-of-the-art MBFL techniques. Our empirical results demonstrate that shifting from correlation to causal responsibility fundamentally improves FL accuracy. Specifically, \tool{} improves \textit{Mean Reciprocal Rank} (MRR) by $16\%$ and successfully locates 31 more bugs in the Top-10 position compared to traditional baselines like \texttt{BLMu}. 
In summary, this paper makes the following contributions:
\begin{enumerate}
    \item \textbf{Theoretical Framework:} We formalize CBFL -- an extension of MBFL as an actual causality problem.
    \item \textbf{Heuristics:} We introduce heuristics to compute the approximate degree of responsibility for faults using distance weights and P2F penalization.
    \item \textbf{Algorithm:} We present an algorithm and its implementation in a tool \tool, a scalable, adaptive, hierarchical, iterative search-based algorithm that generates mutants to approximate actual causality.
    \item \textbf{Empirical Evaluation:} We provide a comprehensive evaluation of \tool on real-world Java bugs demonstrating the superiority of our approach over traditional statistical suspiciousness metrics.
\end{enumerate}

\mypara{Availability}
All \tool's code and experimental results are available at \cite{zenodo:softrex}.

\section{Background}

In this section, we provide a brief, simplified overview of actual causality and MBFL. For the full exposition of actual causality, see Appendix A.

\subsection{Actual Causality}
\begin{figure}[t!]
\centering
    \begin{tikzpicture}[outer sep=auto]
    \node (v1) at (-3, 0) [draw, circle, rounded corners, minimum size=0.5cm,fill=red!10] {$s_1$};
    \node (v2) at (-1.5, 0) [draw, circle, rounded corners, minimum size=0.5cm,fill=red!10] {$s_2$};
    \node (v3) at (0, 0) [draw, circle, rounded corners, minimum size=0.5cm,fill=red!10] {$s_3$};
    \node (dots) at (1.5, 0) {$\mathbf{\cdots}$};
    \node (vn) at (3.0, 0) [draw, circle, rounded corners, minimum size=0.5cm,fill=red!10] {$s_n$};

    \node (f) at (0, -1.0) [draw, rectangle, rounded corners, fill=blue!10, minimum size=0.6cm]{compute the outcome $O$ given the input variables $s_1, s_2,\dots,s_{n}$. };

    \draw [-Triangle] (v1) -- (f);
    \draw [-Triangle] (v2) -- (f);
    \draw [-Triangle] (v3) -- (f);
    \draw [-Triangle] (vn) -- (f);
    
    \node (o) at (0, -2.0) [draw, circle, minimum size=0.5cm, fill=red!10] {$O$};

    \draw [-Triangle] (f) -- (o);
    \end{tikzpicture}
    \caption{A representation of a depth-$2$ causal model used to identify the effect of input variables $s_1, s_2,\dots,s_{n}$ on outcome $O$. In our setting, each $s_i$ is a code element in the program and $O$ an enumeration of execution outcomes. } 
    \label{fig:causal_model}
\end{figure}
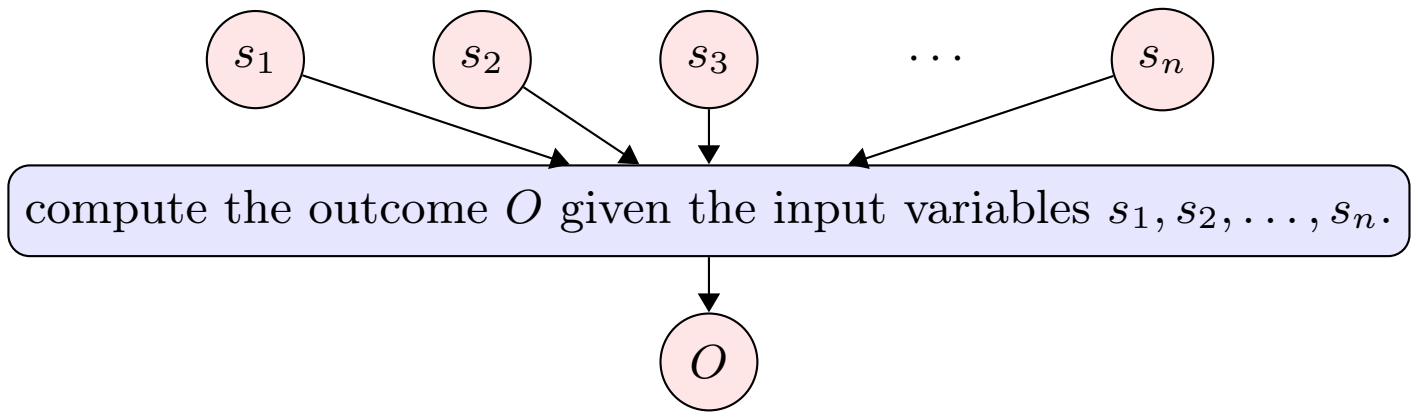

Actual causality~\cite{HP01b,halpern_actual_2016} provides a formal framework for reasoning about the causes of observed outcomes. It differs from \emph{type causality}, formalized by Pearl \cite{pearl_causality_2009}, which deduces general causal relationships between variables from existing data and is forward-looking. In contrast, the study of actual causality is backward-looking, \emph{``what caused a certain test to fail?''}, which is more suitable for analyzing faults in software. The framework describes the world in terms of variables and their values. A causal model $CM$ consists of a set of variables $\V$ and a set of structural equations that capture dependencies between these variables. 
Many applications of actual causality~\cite{navaratnarajah20253d,kelly2023explain} use depth-$2$ causal models, in which a set of independent input variables determines an outcome variable, see \cref{fig:causal_model}. We adopt this approach as well. In these models, causal relationships are examined through interventions that assign alternative values to selected input variables and assess their effects on the outcome.

Informally, an \emph{actual cause} of an observed outcome in a depth-$2$ causal model is a minimal subset of input variables whose values, if changed, alter the outcome.  Following~\cite{Hal15}, we refer to parts of causes as causes. We use the definition of \emph{degree of responsibility}, first introduced in~\cite{chockler2004responsibility}, which quantifies causal influence. Informally, the responsibility of an input variable $X$ for outcome $O$ is $1 / \mid \vec{X} \mid$,
where $\vec{X}$ is a smallest actual cause containing $X$. If there is no actual cause containing $X$, the responsibility of $X$ is $0$. We refer to this subset, $\vec{X} - \{X\}$, as the \emph{contingency} of $X$ and denote its size as $k$, that is, $k = |\vec{X} - \{X\}|$.

\subsection{Mutation-Based Fault Localization (MBFL)}
MBFL relies on creating changes to the source code---known as mutants---which can be analyzed by observing the changes in test execution behavior to expose potential faults in the program. 
Formally, let a buggy target program $\mathcal{P}$ be represented by a finite set of discrete code elements: \(S=\{s_1, s_2,\dots,s_{n}\}\), where each code element has a location and can be of any size. 
The granularity of these elements, ranging from variable expressions and abstract syntax tree nodes to full statements, defines the spatial resolution of the fault localization process, as the final responsibility and suspiciousness scores are attributed directly to these elements in $S$.

To identify the program elements responsible for the failure, MBFL techniques systematically apply mutations to the elements to produce a set of program variants or \textit{mutants} denoted by $M$. We formalize this process via a mapping function $\ell : M \rightarrow 2^S$, which maps each mutant $m \in M$ to the non-empty subset of code elements it modifies. This formalization allows for mutation strategies where a single mutant alters multiple locations simultaneously, \ie{} $|\ell (m)| \geq 1$. Per code element $s \in S$, we define $M_s \subset M$ as the set of all mutants modifying $s$: \[ M_s  = \{ m \in M  | s \in \ell(m) \}. \] 
By evaluating the outcomes of test executions of the mutants in $M_s$, we can systematically measure the impact of $s$ on the overall behavior of $P$. 

The execution behavior of $\mathcal{P}$ and its mutants is evaluated against a test suite $T = F \cup P$, where $F$ is the set of failing tests and $P$ is the set of passing tests. For any test $t \in T$ and program variant $\mathcal{P}'$, let $out(\mathcal{P}', t) \in \{$\emph{pass, fail, exc, timeout}$\}$ denote the execution outcome. This allows us to observe discrete shifts in test execution trajectories, such as transitioning from the original failure to a passing state (\emph{pass}), a process timeout (\emph{timeout}), or a new exception.

To evaluate the suspiciousness of a code element, traditional MBFL constructs a statistical ranking based on execution outcomes. Following the established notion by Wang et al.~\cite{wang_systematic_2025}, for a given $m \in M_s$ applied at location $s$, we define the following transitions:
\begin{itemize}[leftmargin=*]
    \item $f(m)$: the number of tests that \textbf{originally failed} but \textbf{now pass} on $m$.
    \item $p(m)$: the number of tests that \textbf{originally passed} but \textbf{now failed} on $m$.
    \item $nf(m)$: the number of tests \textbf{originally failed} that \textbf{not affected} by $m$.
    \item$np(m)$: the number of tests \textbf{originally passed} that are \textbf{unaffected} by $m$.
    \item $f'(m)$: the number of tests that \textbf{originally failed} and now the test \textbf{output changes} after applying $m$.
\end{itemize}
Using these frequencies, traditional MBFL formulas compute a statistical suspiciousness score, $\mathrm{sus}(m)$, for each mutant. \autoref{tab:mfbl} presents the formulas of the established baseline methods evaluated in this study, where $\alpha$ adjusts the weighting of the terms. \texttt{BLMu} incorporates $\mu$ that denotes the number of tests that change the outcome of the failing test on any mutant over the number of test that transition from passing to failing. 
\newlength{\baseH}
\setlength{\baseH}{0.62cm} 
\begin{table}[t!]
    \small
    \centering
    \caption{Suspiciousness formulas for fault localization methods (for $m \in M_s$).}
    \label{tab:mfbl}
    \setlength{\tabcolsep}{2pt}
    
    \begin{minipage}[t]{0.45\textwidth}
        \centering
        \begin{tabular}{@{} p{2.2cm} p{3cm} @{}}
            \toprule
            \textbf{Method} & \textbf{Formula} \\
            \midrule
            \begin{minipage}[c][\baseH][c]{2.2cm}Barinel~\cite{abreu2007accuracy,10.1109/ASE.2009.25}\end{minipage} 
            & \begin{minipage}[c][\baseH][c]{3cm}
            $1-\frac{p(m)}{p(m)+f(m)}$
            \end{minipage} \\[1pt]

            \begin{minipage}[c][\baseH][c]{2.2cm}DStar2~\cite{6651713}\end{minipage} 
            & \begin{minipage}[c][\baseH][c]{3cm}
                $\frac{f(m)^2}{p(m)+nf(m)}$
            \end{minipage} \\[1pt]

            \begin{minipage}[c][\baseH][c]{2.2cm}Jaccard~\cite{10.1145/2000791.2000795}\end{minipage} 
            & \begin{minipage}[c][\baseH][c]{3cm}
                $\frac{f(m)}{2f(m)+nf(m)}$
            \end{minipage} \\[1pt]

            \begin{minipage}[c][\baseH][c]{2.2cm}Metallaxis~\cite{papadakisMetallaxisFLMutationbasedFault2015b}\end{minipage} 
            & \begin{minipage}[c][\baseH][c]{3cm}
                $\frac{f'(m)}{\sqrt{|F|(f'(m)+p(m))}}$
            \end{minipage} \\
            \bottomrule
        \end{tabular}
    \end{minipage}%
    \hspace{0.4cm}
    \begin{minipage}[t]{0.5\textwidth}
        \centering
        \begin{tabular}{@{} p{2cm} p{3.725cm} @{}}
            \toprule
            \textbf{Method} & \textbf{Formula} \\
            \midrule
            \begin{minipage}[c][0.74\baseH][c]{2cm}BLMu\cite{wang_systematic_2025}\end{minipage} 
            & \begin{minipage}[c][0.74\baseH][c]{3.725cm}
                $\alpha \cdot f(m) + \beta f'(m) - \mu \cdot p(m)$
            \end{minipage} \\[1pt]

            \begin{minipage}[c][0.74\baseH][c]{2cm}Muse~\cite{moon_ask_2014}\end{minipage} 
            & \begin{minipage}[c][0.74\baseH][c]{3.725cm}
                $f(m)-\alpha \cdot p(m)$
            \end{minipage} \\[1pt]

            \begin{minipage}[c][0.74\baseH][c]{2cm}Op2 \cite{10.1145/2000791.2000795}\end{minipage} 
            & \begin{minipage}[c][0.74\baseH][c]{3.725cm}
                $\frac{f(m)-p(m)}{p(m)+f(m)+1}$
            \end{minipage} \\[1pt]
            
            \begin{minipage}[c][1.78\baseH][c]{2cm}Tarantula~\cite{jones2005empirical}\end{minipage} 
            & \begin{minipage}[c][1.78\baseH][c]{3.725cm}
                $\displaystyle
                \frac{
                    \frac{f(m)}{f(m)+nf(m)}
                }{
                    \frac{f(m)}{f(m)+nf(m)}
                    +
                    \frac{p(m)}{p(m)+np(m)}
                }$%
            \end{minipage} \\
            \bottomrule
        \end{tabular}
    \end{minipage}
\end{table}

As multiple mutants can be generated at a single location $s$, MBFL methods aggregate them by calculating the average over all mutant scores. The final suspiciousness ranking for code element $s$ is therefore computed as: \[ \mathrm{Susp}(s)= \frac{1}{|M_s|}\sum_{m \in M_s} sus(m)\] Traditional MBFL methods fail to distinguish between a mutant that leaves a failure completely unchanged and one that completely changes the execution path or the type of failure.


\section{Theoretical Foundation of Causality-Based Fault Localization}
\label{sec:FL:foundaitons}
To move beyond correlation, we introduce the use of actual causality for FL. The goal is to determine whether a code element $s$ is an actual cause of a test $f \in F$ failing. Specifically, we define a causal model $CM_f$ for a failed test $f$, where its input variables correspond to the code elements within the coverage of $f$. In this formulation, interventions on these variables are equivalent to mutations. In addition, $CM_f$ includes an additional variable, $O$, corresponding to the execution outcome of the---possibly intervened/mutated---program. This results in a depth-$2$ causal model (\autoref{fig:causal_model}), where each code element is causally independent, meaning that each element can be mutated without altering others. 

Informally, a code element $s$ is a cause of a test failure $f \in F$ if an intervention on $s$, applied via some mutation $m$, alters the test's execution behavior. However, in real-world settings, single interventions rarely suffice; faults are often masked by secondary errors and may require simultaneous modifications across the program to be revealed. The benefits of an explicit causal model allow evaluating under \emph{contingencies}---sets of other code elements that are intervened upon to reveal the causal effect of $s$.


Unlike traditional approaches that only look for a complete repair ($\mathrm{fail} \rightarrow \mathrm{pass}$), we consider any deviation from the original failure trajectory to be potential evidence of a causal link. We formalize this using an indicator of outcome for a failing test $f\in F$ and mutation $m$: \[
    R(m,f)=\mathbf{1}\{out(\mathcal{P},f)=\mathrm{fail}\ \wedge\ out(m, f) \neq out(\mathcal{P},f) \}.
  \]
Thus, for a code element $s$ to have a responsibility greater than $0$ for the failed test $f$, our causal model requires that there exists some mutant $m \in M$ such that $s \in \ell(m)$ and $R(m, f) = 1$.

\subsection{Practical Causal Relaxation for Software}
While formally rigorous, computing exact actual causes across arbitrary programs is computationally intractable due to the size of the space of multi-point code interventions. To make causal FL practical, we approximate the degree of responsibility, as defined in \autoref{def:resp}, by combining the following: encoding behavior shifts, representative mutant sampling, side-effect penalization, and interventional cost.

\mypara{Outcome Sensitive Responsibility} Rather than exhaustively searching for a minimal contingency size $k$, we estimate the responsibility of an intervention for a failing test $f \in F$ by considering the outcome change. Specifically, in the case that there is a partial shift in the failing test's trajectory, such as a modified stack trace, a different exception type, or a process timeout. This signals that $m$ perturbs the causal chain of the bug even if it does not resolve it: $out(m, f) \neq out(\mathcal{P}, f) \wedge out(m, f) \neq \mathrm{pass}$, where $m$ is the mutation applied at the location $s$. We scale the responsibility so that in these scenarios the code element $s$ can be at most $0.5$. For example, in a scenario where a mutant $m$ alters the execution behavior of a failing test without achieving full repair, but does so by requiring a minimum of $k=3$ concurrent interventions, the assigned responsibility scales accordingly. We evaluate the resulting responsibility at $s$ using the upper estimate: \[\mathrm{Resp}_f(s) = \frac{0.5}{0.5+k}=\frac{0.5}{0.5+3} = \frac{0.5}{3.5} \approx 0.143.\] However, if $k=0$ and there is partial shift, we assign responsibility of 0.5 to $s$. As $k$ must denote the smallest number of additional interventions needed to induce this shift, this example illustrates how the framework, by default, scales to penalize highly entangled, multi-point causal dependencies.

\mypara{Mutant Sampling} Evaluating every possible multi-code intervention across a whole program is computationally intractable.\footnote{For a program of size $n$, there are $O(2^n)$ subsets of possible intervention locations to consider.} Instead, we estimate the responsibility of a code element $s \in S$ over an evaluated sample of representative mutations, $M_s$. Our causal responsibility score is agnostic to how $M_s$ is generated; it remains fully computable given an arbitrary set of mutants, including the exhaustive single-point mutations generated by tools like Major~\cite{major}.

While sampling sacrifices exact theoretical completeness, it yields a practical approximation of the causal responsibility of $s$. If no mutant in $M_s$ alters the failure trajectory, we cannot definitively rule out that a more complex, undiscovered multi-point intervention might do so. However, if any mutant $M_s$ successfully triggers a behavior shift, the causal link exists and the element's true responsibility is guaranteed to be at least the value computed.

Therefore, the accuracy of the approximation to the actual causal responsibility relies on the diversity of the mutation operator applied. By utilizing a comprehensive suite of operators---such as statement deletions, literal value replacement, and relational and conditional operator replacements---we maximize the interventional coverage of the code element, ensuring a high-quality, practical approximation of the true causal responsibility of $s$.

\mypara{Pass-to-Fail (P2F) Penalization} A fault-repairing intervention should not break non-faulty program behavior captured by the test suite. If an intervention $m$ alters the outcomes of originally passing tests $P$, its responsibility is scaled by a validity ratio :
\begin{equation}\label{eq:valid}
    V(m) = 1.0 - \frac{|\{t \in P \mid out(m, t) \neq \mathrm{pass}\}|}{|P|}
\end{equation} This penalizes destructive mutations that induce widespread incorrect behavior. Incorporating passing-test behavior is a foundational axiom in traditional MBFL. \muse{}, for example, subtracts the number of mutants that have a passing-to-failing test transitions for a given code element, whereas Metallaxis incorporates this into its denominator. Our framework explicitly isolates this as a credibility bound on a causal claim, using only the relevant mutant rather than all mutants for code element $s$. This addresses how to deal with common mutants that satisfy execution change of the failing test, however change behavior of existing logic, preserving the distinction between true root-cause repair and coincidental anomaly.

\mypara{Distance Weighting} To penalize complex, large or disproportionate interventions, we regularize the responsibility with an intervention weight $dist(m,s)$. Massive alterations may change the test outcome simply by removing an execution path rather than exposing the specific fault. The distance-based weighting function captures either the syntactical structural change or semantic embedding distance (\eg{} CodeBERT representational shift) between the original code at location $s$ and mutant $m$. For the structural distance heuristic, the cost scales according to the size of the target code element $s$, see \Cref{sec:conf} for specific details. Alternatively, the semantic distance quantifies the representational shift between the mutated and original code embeddings, penalizing semantically destructive intervention regardless of the physical size.

The final ranked score for a code element $s$ for a failing test $f$ is the maximum weighted responsibility across its mutants:

\begin{equation}\label{eq:softrex}
    \mathrm{Rank}^{*}_f(s) = \max_{m \in M_s}\left( \frac{\mathrm{Resp}_f(m) \cdot V(m)}{1 + \lambda \cdot \mathrm{dist}(s, m)}\right),
\end{equation}
where $\lambda > 0$ is a hyperparameter balancing causal responsibility against intervention complexity. Each failing test represents a distinct causal scenario; we aggregate these individual location rankings across all the failing tests to produce a global FL score, enabling direct evaluation against standard correlation based MBFL techniques.



%

\section{\tool Algorithm}
\label{sec:softrex:alg}
%
In this section, we introduce the \tool{} search algorithm, which operationalizes our theoretical framework by dynamically generating causal interventions. Computing the exact minimal contingency set $k$ across an entire program is an $O(2^{|S|})$ combinatorial problem. Further, the space of syntactical mutations for any single code element can be extremely large; an exact exhaustive search is computationally unfeasible. To address this, \tool{} employs a hierarchical, iterative, adaptive syntax-tree-based search that isolates the smallest blocks of the program without requiring an exhaustive search.

\begin{algorithm}[t!]
\caption{\tool{} Search} 
\label{algo:cbflsearch}
\begin{algorithmic}[1]
\Require Program $\mathcal{P}$, failing test $f$, test suite $T$, trials per group $n$, min region size $s$
\Ensure Mutant/test outcome matrix $A$.

\State $B \gets [\,]$ \Comment{Initialize baseline results to empty}
\State $\forall t \in T : B[t] \gets out(\mathcal{P}, t)$ \Comment{Record baseline test outcomes}
\State $\mathcal{G} \gets \text{InitialGroups}(\mathcal{P}.\text{CST})$ \Comment{Initial groups of mutable CST blocks}
\State $A \gets [\,]$

\For{$G \in \mathcal{G}$}
    \State $\mathcal{M} \gets \text{GenBatches}(G, n)$
    \For{$m \in \mathcal{M}$}
        \State $\mathcal{P}' \gets apply(\mathcal{P},m)$ \Comment{Replace block with type-compatible simple value}
        \If{\label{line:cannot-be-evaluated}$\mathcal{P}'$ \text{cannot be evaluated}}
            \State \textbf{continue}
        \EndIf
        
        \For{\label{line:test-all-start}$t \in T$}
            \State $o_t \gets out(\mathcal{P}', t)$ \Comment{Evaluate each test in the suite on the mutant}
            \State\label{line:test-all-end} $A_{m,t} \gets (o_t = B[t])$ \Comment{Record comparison in outcome matrix}
        \EndFor
    
        \If{$out(\mathcal{P}', f) = \text{pass} \lor out(\mathcal{P}', f) \neq B[f]$}
            \If{$(|G| > s)$}
                \State \label{line:subdivide} $\mathcal{G} \gets \mathcal{G} \cup \text{Subdivide}(G)$ \Comment{Recursively subdivide block group}
            \EndIf
        \EndIf
    \EndFor
\EndFor

\State \Return $A$
\end{algorithmic}
\end{algorithm}

\Cref{algo:cbflsearch} accepts a failing test $f$, which drives the search alongside a broader test suite $T$. First, \tool{} records the baseline test outcomes $B$ and restricts the search to source files covered by the driving test $f$, using per-test coverage report. The covered files are then searched independently: for each file, \tool{} parses the source into a CST and partitions the file into coarse groups of mutable CST blocks ($\mathcal{G}$). For each group, we generate batches of mutated variations (up to $n$ trials per group) by replacing the targeted block with a type-compatible simple value, $apply(\mathcal{P}, m)$---for example, substituting a block whose return type is \textsc{int} with $1$, $0$, or \textsc{MAX\_INT}. If a mutation yields an uncompilable or malformed variant that cannot be evaluated, it is discarded (line~\ref{line:cannot-be-evaluated}). This filtering step ensures that only valid, executable interventions are considered in our analysis.
For each valid mutant, we evaluate the entire test suite to identify changes in either the success of the individual tests or the specific errors they return (lines \ref{line:test-all-start}--\ref{line:test-all-end}). If the outcome of the driving test $f$ changes, it indicates that the mutated region warrants further exploration. Consequently, provided the block size exceeds the minimum threshold ($|G| > s$), we subdivide that block and recursively continue the search (line~\ref{line:subdivide}).
Ultimately, these evaluations construct the outcome matrix $A$. This matrix directly provides the execution states necessary to compute both the passing-test validity penalty and the base responsibility defined in~\Cref{sec:FL:foundaitons}.
\section{Evaluation Methodology}
\label{sec:eval}
To assess our causality-based framework's effectiveness and the \tool{} search algorithm's practical viability, we conduct a comprehensive empirical evaluation. Our study is designed to answer the following Research Questions (RQs): 
\begin{enumerate}[label=\textbf{RQ\arabic*.}, ref=RQ\arabic*, leftmargin=*]
\label{sec:rqs}
    \item \label{RQ1: Locality} How does \tool's causality-based ranking formula compare against state-of-the-art MBFL formulas when using a fixed mutant sample?
    \item \label{RQ2: Ablation Study} To what extent do (a)~penalty regularization and (b)~structural/semantic distance regularization individually improve causal ranking accuracy?
    \item \label{RQ3: Search Impact} To what extent does guiding the mutation generation improve FL accuracy and efficiency over standard mutation strategies?
\end{enumerate}

\subsection{Experiment Design}
\label{sec:exp:design}

We structure our evaluation into two distinct experiments. 

\mypara{Experiment 1} We isolate the effectiveness of \tool{}'s causal ranking formula from its heuristic mutation search algorithm. 
We replace \tool{}'s search with \textsc{Major}~\cite{major} to systematically generate single-line mutants via mutation analysis across the target program by applying our \tool's scoring to this static mutant matrix. This allows for a direct comparison against existing MBFL formulas. 

\noindent \textbf{Experiment 2} focuses on evaluating the whole pipeline, integrating our hierarchical search (\Cref{algo:cbflsearch}). This experiment assesses whether the targeted sampling of mutations can achieve effective fault localization without the immense cost of exhaustive mutation testing.

\subsection{Configurations and Metrics}
\label{sec:conf}
\mypara{Configurations}
To answer our RQs, we evaluate \SoAno{} state-of-the-art (SoA) baselines and 6 configurations of \tool, and additional hyperparameter variations for \tool{} with $\lambda \in [0.05, 0.1, 0.25, 0.5, 1, 2, 3]$. For the 8 baseline methods, we re-implemented the standard MBFL scoring formulas as outlined in~\autoref{tab:mfbl}. These include established MBFL approaches, namely \textsc{Metallaxis}~\cite{papadakisMetallaxisFLMutationbasedFault2015b}, \textsc{BLMu}~\cite{wang_systematic_2025} and \textsc{Muse}~\cite{moon_ask_2014}, alongside SBFL formulas mapped to different mutation outcomes: \textsc{Tarantula}~\cite{jones2005empirical}, \textsc{Barinel}~\cite{abreu2007accuracy,10.1109/ASE.2009.25}, \textsc{Dstar2}~\cite{6651713}, \textsc{Jaccard}~\cite{Wong2016Survey}, and \textsc{Op2}~\cite{10.1145/2000791.2000795}. We avoid direct comparison to SBFL techniques to ensure a strictly fair comparison.
To rigorously assess our causal framework, we conduct an ablation study across 6 \tool{} configurations for each experiment. We evaluate the foundational causal formula as \rexbase{} that directly aligns the rank to the raw degree of responsibility, $\mathrm{Resp}_f(m)$. To isolate the individual contributions of regularization mechanisms, we evaluate the P2F penalization independently as \rexpenalty{} and the distance weighting independently as \rexdist{}. We use two distance measures: one based on structural cost changes~\rexdist{} and another that incorporates CodeBERT's~\cite{feng2020codebert} cosine embedding distances,~\rexdistsem{}. The structural distance depends on the element type of $s$ and whether the element was deleted, replaced or modified, with further details provided in the appendix. For Experiment 2, where mutations can span multiple lines, we compute this weighting by averaging the distance weights across all affected elements $s$. Finally, the complete \tool{} is represented by the configurations where both P2F penalization and distance weighting mechanisms are fully active (\rexpendistsem{}, \rexpendist{}).



\mypara{Evaluation Metrics}
We compare the rankings produced by \tool against SoA (\autoref{tab:mfbl}) by measuring the \textit{Mean Average Rank} (MAR), \textit{Mean Reciprocal Rank} (MRR), and Top-1, Top-3, Top-5, and Top-10 accuracy. MAR represents the average rank of all bugs per approach. Top-$N$ defines the number of bugs in which the faulty code elements are located at top $N$ position of the ranking given by the approach.

\subsection{Benchmark and Environment}\label{sec:bench}
We utilize \href{https://github.com/rjust/defects4j}{Defects4J}~\cite{just_defects4j_2014} (commit \texttt{8c16da8}), a collection of reproducible bugs in Java programs for which the location of the real fault and the corresponding test triggering the error before the fix are known. 
\emph{Experiment 1} evaluates 827 bugs across 17 projects: JacksonXML (\texttt{JXml}), \texttt{Cli}, JacksonCore (\texttt{JCore}), \texttt{Compress}, \texttt{Jsoup}, \texttt{Codec}, JacksonDatabind (\texttt{JDB}), \texttt{Math}, \texttt{Collections}, \texttt{Csv}, \texttt{JxPath}, \texttt{Gson}, \texttt{Lang}, \texttt{Chart}, \texttt{Time}, \texttt{Closure} and \texttt{Mockito}; 27 bugs were excluded due to Out-of-Memory or Timeout exceptions. 
\emph{Experiment 2} is performed on a representative subset (\texttt{JacksonXml}, \texttt{Gson} \texttt{Csv}, \texttt{Math}) chosen via uniform random sampling. We selected projects spanning different scales and failure densities: \texttt{Math} provides a large, failure dense stress test; \texttt{CSV} is more compact, but contains several non-trivial bugs. \texttt{Gson} and \texttt{JacksonXml} provide sparser serialization focused targets. Full project and execution details are in the Appendix.



Experiments were executed on {\em anonymized for submission} HPC cluster within a 64-bit Linux environment managed by the \emph{Sun Grid Engine} (SGE) scheduler. Jobs were submitted to the generic CPU queue, configured to utilize 4 cores and 32 GB of RAM.
The max timeout for each job was 48 hours, with no experiment reaching it, and~600 seconds per task.

\section{Results and Evaluation}
\subsection{RQ1 - Baseline Comparison}
\label{rq1:ans}

\pgfplotsset{compat=1.18}

\definecolor{colMetallaxis}{HTML}{2A78D6}  
\definecolor{colBarinel}{HTML}{1BAF7A}     
\definecolor{colTarantula}{HTML}{EB6834}   
\definecolor{colJaccard}{HTML}{EDA100}     
\definecolor{colDstar2}{HTML}{E87BA4}      
\definecolor{colMuse}{HTML}{6B7280}        
\definecolor{colOp2}{HTML}{7C3AED}         
\definecolor{colSoftrex}{HTML}{B91C1C}     
\definecolor{colBLMu}{HTML}{15803D}        

\begin{figure}[t!]
\begin{tikzpicture}
\begin{axis}[
    width=\textwidth,
    height=4cm,
    ybar,
    bar width=5pt,
    ylabel={Number of bugs located},
    symbolic x coords={Top-1, Top-3, Top-5, Top-10},
    xtick=data,
    xticklabel style={font=\small},
    yticklabel style={font=\small},
    ymin=0,
    enlarge x limits=0.15,
    legend style={
        at={(0.5,1.42)},
        anchor=north,
        legend columns=5,
        draw=none,
        fill=none,
        font=\small,
        column sep=6pt,
    },
    ymajorgrids=true,
    grid style={dashed, gray!25},
    tick align=outside,
    xlabel style={font=\small, yshift=-4pt},
    ylabel style={font=\small},
]

\addplot[ybar, fill=colMetallaxis, draw=colMetallaxis]
    coordinates {(Top-1, 40) (Top-3, 93) (Top-5, 122) (Top-10, 185)};
\addlegendentry{Metallaxis}

\addplot[ybar, fill=colBarinel, draw=colBarinel]
    coordinates {(Top-1, 71) (Top-3, 122) (Top-5, 140) (Top-10, 176)};
\addlegendentry{Barinel}

\addplot[ybar, fill=colTarantula, draw=colTarantula]
    coordinates {(Top-1, 67) (Top-3, 113) (Top-5, 132) (Top-10, 171)};
\addlegendentry{Tarantula}

\addplot[ybar, fill=colJaccard, draw=colJaccard]
    coordinates {(Top-1, 59) (Top-3, 113) (Top-5, 131) (Top-10, 170)};
\addlegendentry{Jaccard}

\addplot[ybar, fill=colDstar2, draw=colDstar2]
    coordinates {(Top-1, 63) (Top-3, 123) (Top-5, 140) (Top-10, 176)};
\addlegendentry{Dstar2}

\addplot[ybar, fill=colMuse, draw=colMuse]
    coordinates {(Top-1, 68) (Top-3, 120) (Top-5, 139) (Top-10, 187)};
\addlegendentry{Muse}

\addplot[ybar, fill=colOp2, draw=colOp2]
    coordinates {(Top-1, 71) (Top-3, 122) (Top-5, 140) (Top-10, 176)};
\addlegendentry{Op2}

\addplot[ybar, fill=colBLMu, draw=colBLMu]
    coordinates {(Top-1, 89) (Top-3, 168) (Top-5, 211) (Top-10, 270)};
\addlegendentry{BLMu}

\addplot[ybar, fill=colSoftrex, draw=colSoftrex]
    coordinates {(Top-1, 103) (Top-3, 203) (Top-5, 244) (Top-10, 301)};
\addlegendentry{\tool{}}
\end{axis}
\end{tikzpicture}
\caption{Top-$N$ bugs located with different thresholds aggregated over all projects evaluated ($N \in \{1,3,5,10\}$).}
\label{fig:globaltopN}
\end{figure}
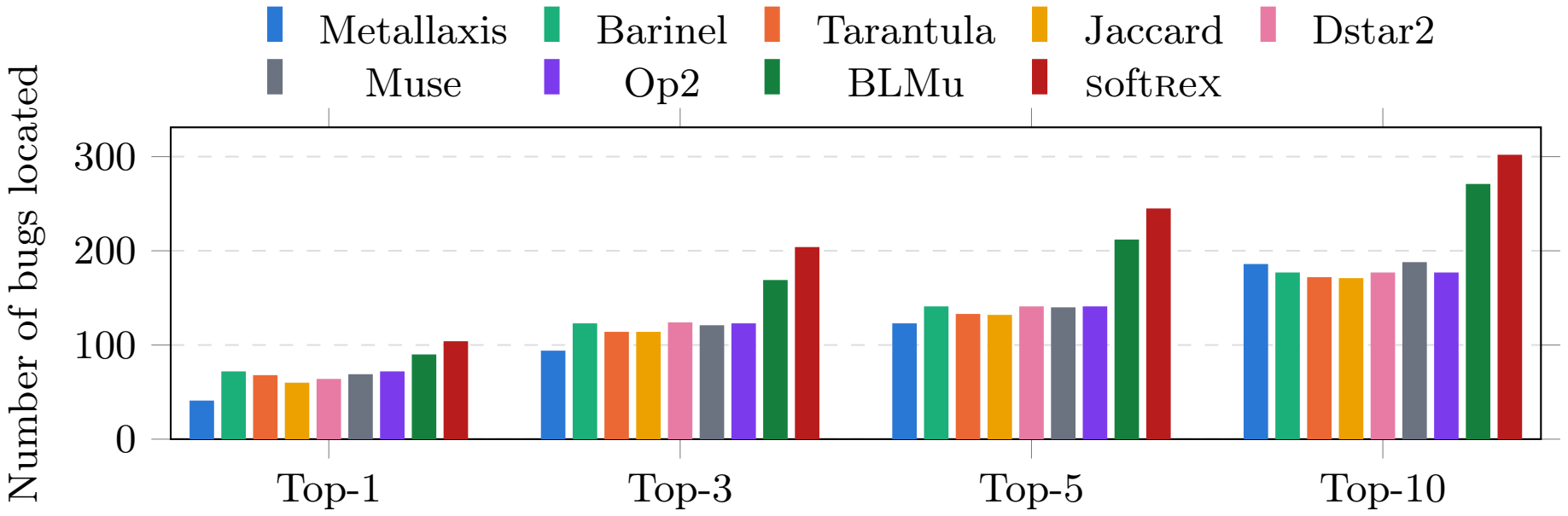

We evaluate \tool{} against \SoAno{} 
established baseline methods over a static mutant matrix. 
\autoref{fig:globaltopN} presents aggregated global FL performance across all projects. 
Our \tool{} with the specific configuration~\rexpendistsem{}, where $\lambda=0.05$ outperforms all baselines across every ranking threshold. We successfully isolate 103 bugs at exactly Top-1, outperforming the strongest MBFL method, BLMu, with 89 bugs. Further, the performance gap widens as $N$ increases, with \tool{} finding 301 bugs in the Top-10 compared to \texttt{BLMu}'s 270, demonstrating a higher overall reliability and less noise. \tool{} achieves a global MRR of 0.349, establishing a clear margin over traditional methods. \texttt{BLMu} and \texttt{Barinel} have MRRs of 0.301 and 0.222, respectively, while \texttt{Op2} severely underperforms with 0.113. The relative improvement in MRR ($\approx 15\%$) demonstrates that the causal-based method provides a less noisy ranking. While our \tool{} noticeably improves Top-N and MRR, we observe a marginal trade-off in MAR, where 
BLMu
scores slightly better (45.58 vs. \tool{}'s 48.13).

\mypara{Project-Level Analysis}
We further break down the performance across the 17 Defects4j projects. Compared with the best performing baseline MBFL technique per project, \tool{} 
improves average MRR by $+0.0670$. Examining the number of bugs located at various ranking thresholds (Top-N) reveals \tool{}'s consistent superiority across diverse project domains. Naturally, projects with a larger number of evaluated bugs, such as \texttt{Math} and \texttt{Closure}, yield higher Top-N counts across all evaluated methods. However, \tool{} has an advantage over SoA, irrespective of project scales. We observe only a few isolated cases where a traditional MBFL approach outperforms \tool on MRR: 
\texttt{Cli} and \texttt{JacksonXml}. 

To verify the robustness of the final \cref{eq:softrex}, we conducted a sensitivity analysis on the weight parameter $\lambda$ detailed in the Appendix. The results indicate \tool{}'s stability across a wide range of values while avoiding overfitting, whereas \texttt{BLMu}'s hyperparameter drastically change the results entirely. We used $\lambda = 0.05$ for our evaluation as it empirically provides the optimal balance.

\subsection{RQ2 - Ablation Study of \tool}
\label{rq2:ans}
To understand how individual components contribute to \tool's performance, we evaluate 4 structural ablations of the proposed ~\cref{eq:softrex}: the base responsibility (\rexbase{}), the integration of the P2F penalization, the distance weighting, and the fully combined formula as shown in~\cref{eq:softrex}. 

\mypara{Base Causal Responsibility} The base responsibility configuration achieves an average MRR of 0.269, beating the other approaches and placing it just below \texttt{BLMu}. Beyond MRR, the causal formulation shows its strength: \rexbase{} located 284 bugs in the Top-10 and achieved an overall MAR of 45.903. In both these metrics, ~\rexbase{} outperforms every traditional baseline evaluated, except \texttt{BLMu} on MAR. This shows that computing pure causal responsibility is very effective at isolating the general faulty neighborhood.

\mypara{P2F penalization} Applying the penalization (\rexpenalty{}) yields a significant performance leap, increasing MRR to 0.335. This supports our hypothesis that penalizing mutants that break previously passing tests is crucial for filtering out \textit{pseudo-fixes} that satisfy the change in observed behavior by reducing the program's overall logic. This filtering effect is very visible in the Top-N metric as Top-1 increases by 50\% (from 50 to 75 bugs) and Top-10 peaks at 311 bugs over all configurations. This shows the constraint pushes the true root cause to the top of the ranking.

\mypara{Distance Weighting} Utilizing both heuristic and semantic distance weighting independently also strengthens the base responsibility calculation. Specifically, CodeBERT semantic distance pushes Top-1 accuracy to 93. The heuristic distance does not improve to the same magnitude as semantic distance, yet has the lowest (best) MAR out of the evaluated methods. The base responsibility often assigns the exact same responsibility scores to multiple statements, and an effective tie-breaking mechanism is required. The distance measurements resolve ties by favoring minimal shifts over massive changes, achieving high precision even without considering any outcomes from passing tests.

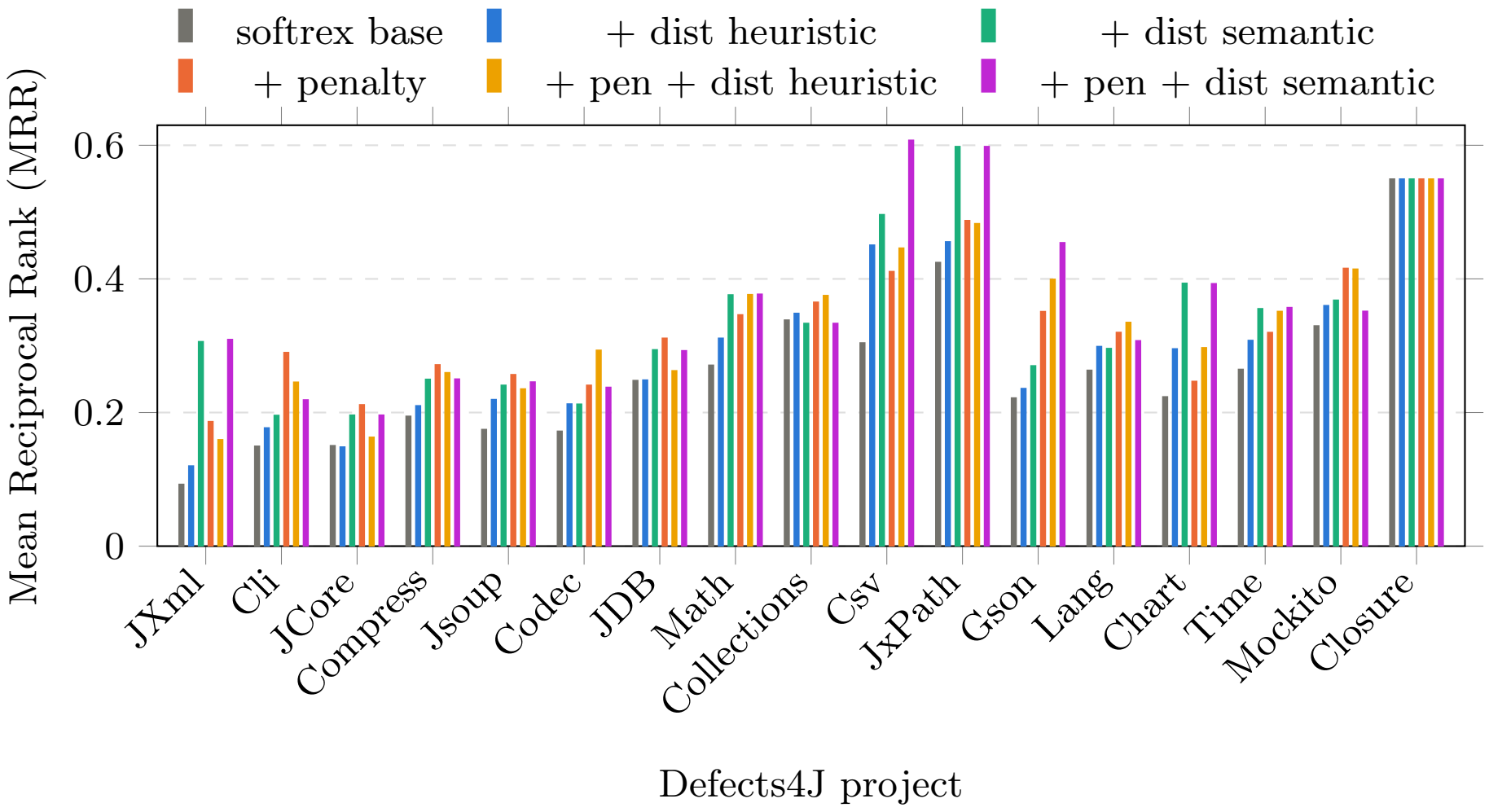
\begin{figure}[t!]    
    \centering
\begin{tikzpicture}
\begin{axis}[
    width=\textwidth,
    height=5cm,
    ybar=1.25pt,
    bar width=1pt,
    xlabel={Defects4J project},
    ylabel={Mean Reciprocal Rank (MRR)},
    symbolic x coords={
        JXml,Cli, JCore,Compress, Jsoup, Codec,JDB, Math, Collections,Csv,JxPath, Gson,Lang,Chart,Time, Mockito,Closure
    },
    xtick=data,
    xticklabel style={rotate=45, anchor=east, font=\small},
    yticklabel style={font=\small},
    ymin=0, ymax=0.63,
    enlarge x limits=0.04,
    legend style={
        at={(0.5,1.3)},
        anchor=north,
        legend columns=3,
        draw=none,
        fill=none,
        font=\small,
        column sep=8pt,
    },
    ymajorgrids=true,
    grid style={dashed, gray!25},
    tick align=outside,
    xlabel style={font=\small, yshift=-4pt},
    ylabel style={font=\small},
]

\addplot[ybar, fill=colBase, draw=colBase] coordinates {
(Closure, 0.5492) (Math, 0.2705) (JDB, 0.2476) (Lang, 0.2629) (Time, 0.2643) 
(Mockito, 0.3294) (JxPath, 0.4243) (Collections, 0.3381) (Jsoup, 0.1743) 
(Csv, 0.3039) (Chart, 0.2232) (Compress, 0.1941) (JCore, 0.1501) 
(Gson, 0.2215) (Codec, 0.1717) (Cli, 0.1493) (JXml, 0.0922) 
};
\addlegendentry{softrex base}

\addplot[ybar, fill=colDistH, draw=colDistH] coordinates {
(Closure, 0.5492) (Math, 0.3109) (Jsoup, 0.2191) (JDB, 0.2483) (Lang, 0.2985) 
(Time, 0.3076) (Csv, 0.4503) (Mockito, 0.3597) (JxPath, 0.4552) 
(Collections, 0.3481) (Chart, 0.2950) (Compress, 0.2098) (Cli, 0.1767) 
(JCore, 0.1481) (Gson, 0.2356) (Codec, 0.2125) (JXml, 0.1197) 
};
\addlegendentry{$+$ dist heuristic}

\addplot[ybar, fill=colDistS, draw=colDistS] coordinates {
(Closure, 0.5492) (Math, 0.3758) (JDB, 0.2936) (Jsoup, 0.2405) (Lang, 0.2955) 
(JxPath, 0.5978) (Chart, 0.3931) (Time, 0.3551) (Compress, 0.2494) 
(Csv, 0.4959) (Mockito, 0.3678) (Collections, 0.3331) (JCore, 0.1958) 
(JXml, 0.3058) (Codec, 0.2122) (Cli, 0.1954) (Gson, 0.2696) 
};
\addlegendentry{$+$ dist semantic}

\addplot[ybar, fill=colPen, draw=colPen] coordinates {
(Closure, 0.5492) (Math, 0.3459) (JDB, 0.3109) (Jsoup, 0.2564) (Mockito, 0.4155) 
(Lang, 0.3196) (Time, 0.3195) (JxPath, 0.4870) (Compress, 0.2711) 
(Collections, 0.3648) (Cli, 0.2895) (JCore, 0.2113) (Csv, 0.4107) 
(Chart, 0.2463) (Gson, 0.3508) (Codec, 0.2405) (JXml, 0.1861)
};
\addlegendentry{$+$ penalty}

\addplot[ybar, fill=colPenDistH, draw=colPenDistH] coordinates {
(Closure, 0.5492) (Math, 0.3762) (Jsoup, 0.2350) (JDB, 0.2622) (Lang, 0.3346) 
(Mockito, 0.4143) (Time, 0.3510) (Compress, 0.2593) (JxPath, 0.4824) 
(Csv, 0.4458) (Collections, 0.3748) (Chart, 0.2967) (Cli, 0.2451) 
(Codec, 0.2929) (JCore, 0.1627) (Gson, 0.3991) (JXml, 0.1590) 

};
\addlegendentry{$+$ pen $+$ dist heuristic}

\addplot[ybar, fill=colPenDistS, draw=colPenDistS] coordinates {
(Math, 0.3768) (Closure, 0.5492) (JDB, 0.2922) (Jsoup, 0.2454) (Lang, 0.3071) 
(JxPath, 0.5978) (Csv, 0.6072) (Chart, 0.3925) (Time, 0.3568) 
(Compress, 0.2498) (Mockito, 0.3512) (Collections, 0.3331) (Codec, 0.2374) 
(Cli, 0.2186) (JCore, 0.1958) (Gson, 0.4539) (JXml, 0.3090) 
};
\addlegendentry{$+$ pen $+$ dist semantic}

\end{axis}
\end{tikzpicture}
    \caption{Ablation Study: Mean Reciprocal Rank (MRR) for each project evaluated.}
    \label{fig:projectmrrablation}
\end{figure}
\mypara{Combined: \tool{}} The fully combined configuration achieves best Top-1 and MRR performance across our evaluations, successfully locating 103 bugs at Top-1 and reaching an MRR of 0.349 at $\lambda =0.05$. This shows that our weighting mechanisms are orthogonal and work well together. The validity penalty acts as a filter to eliminate any fake fixes, while the distance weighting ensures the true root cause is pushed to the top of the ranking. Furthermore, substituting the basic syntactic cost heuristic (\rexpendist{}) with deep semantic embeddings (\rexpendistsem{}) yields a 27\% relative increase in Top-1 accuracy (from 81 to 103 bugs), confirming that modeling semantic intent is far superior to simple structural change cost when resolving ties in the ranking.

\mypara{Project-Level Analysis}~\autoref{fig:projectmrrablation} breaks down the ablation study at the project level with respect to MRR. \textit{softrex base} configuration consistently establishes the lower bound for MRR across nearly all 17 projects. Applying the P2F penalization or the semantic distance weighting isolation produces significant but project-dependent improvements. For example, the P2F penalization yields notable gains over \texttt{Mockito}, \texttt{Compress}, \texttt{Cli} and \texttt{JDB}, and semantic distance independently dominates on projects like \texttt{JxPath} and \texttt{JXml}. The fully combined configuration shows the best stability; this is especially visible in projects like \texttt{Gson} and \texttt{Csv}. Lastly, the project-wise breakdown confirms that semantic embeddings consistently provide substantially stronger, more reliable ranking signals than basic syntactic structural costs across the majority of the Defects4J dataset.

\subsection{RQ3 - Guided Mutation Generation Effectiveness}
\label{rq3:ans}
While \textit{Experiment 1} establishes the theoretical superiority of our ranking formula under a static mutant pool, \textit{Experiment 2} evaluates the end-to-end \tool{} pipeline utilizing our hierarchical search algorithm. 
We evaluate each approach over a uniform random sample of bug IDs from 4 representative projects.
\begin{table}[t!]
    \centering
    \caption{Project-level comparison on matched bug subsets from Major (Experiment 1.), and \tool search (Experiment 2.).}
    \label{tab:experiment2_matched_comparison}
    \footnotesize
    \renewcommand{\arraystretch}{0.7} 
\begin{tabular}{llrrrrrrr}
\toprule
Project & Experiment & Top-1 & Top-3 & Top-5 & Top-10 & MAR & MRR & Mutants\\
\midrule
\multirow{2}{*}{Csv} 
& Major (1) & 2 & 3 & 5 & 6 & 12.41 & 0.294 & 2743 \\
& \tool (2) & 3 & 4 & 5 & 5 & 82.50 & 0.347 & \textbf{958} \\
\midrule
\multirow{2}{*}{Gson} 
& Major (1) & 1 & 3 & 5 & 5 & 5.42 & 0.482 & 372 \\
& \tool (2) & 0 & 0 & 0 & 0 & 87.00 & 0.018 & \textbf{191} \\
\midrule
\multirow{2}{*}{JacksonXml} 
& Major (1) & 0 & 0 & 2 & 3 & 26.5 & 0.144 & 1423 \\
& \tool (2) & 0 & 0 & 1 & 1 & 514.33 & 0.038 & \textbf{1003} \\
\midrule
\multirow{2}{*}{Math} 
& Major (1) & 8 & 14 & 15 & 17 & 8.18 & 0.226 & 19565 \\
& \tool (2) & 7 & 15 & 17 & 19 & 81.67 & 0.247 & \textbf{7126}\\
\bottomrule
\end{tabular}%

\end{table}  

\autoref{tab:experiment2_matched_comparison} compares performance on matched bug sets. Unlike exhaustive mutation baselines, \tool{} employs a targeted search, reducing evaluated mutants by $\approx 2.6\times$ while preserving competitive localization accuracy.
For instance, in larger codebases like \texttt{Math}, \tool{} exhibits a minor trade-off in Top-1 accuracy (locating 7 bugs to Major's 8), yet it matches or surpasses the exhaustive baseline across higher thresholds (Top-3, Top-5, and Top-10) and achieves a notably higher MRR on both \texttt{Math} and \texttt{Csv}. Conversely, smaller or sparsely populated target sets (such as \texttt{Gson} and \texttt{JacksonXml}) reflect the inherent challenges of dynamic sampling under strict search boundaries. Overall, these findings demonstrate that guided hierarchical mutation generation provides an effective balance between FL precision and computational feasibility. Additionally, across all tested bugs, however, \tool{} search matches or beats all other MBFL approaches in Top-N accuracy, while using on average $2.6\times$ fewer mutants evaluated than Major.

\section{Discussion} 

%

Our causal responsibility is approximated using sampled mutants rather than exhaustive interventions. Our validity ratio and distance weighting introduce modeling assumptions; we mitigate this through a comprehensive ablation study and a hyperparameter sensitivity analysis on $\lambda$, which confirmed performance stability. Finally, we applied standard FL metrics (MAR, MRR, and Top-\(N\)), though alternative metrics could yield different perspectives. Focusing solely on program structure while omitting data flow captures static architectural or control-flow dependencies, but it limits the model's ability to represent runtime values, state propagation, or actual execution. Integrating both inside the SCM has been left to future work. 
We evaluated multiple Defects4J projects, supporting generalization across Java codebases. Although our underlying \tool's is language independent, its generalization to other programming languages remains to be validated in the future.

\mypara{Future Work}%
Beyond the controlled benchmark evaluation presented here, we will study \tool in the wild and on other benchmarks. Systems can include substantial setup logic, low-level implementation code, external libraries, and complex code, creating many program elements that correlate with a failure without necessarily causing it. These settings provide a useful test of whether CBFL addresses better FL. Additionally, we will extend this work and investigate how causal-model granularity affects fault ranking, comparing source-code lines with alternative program representations such as \textit{Abstract Syntax Tree} (AST) nodes and program dependence graph nodes. In particular, a depth-$2$ causal model is likely a simplification of the actual causal structure in a software object. Combining SBFL and causal discovery could lead to significant gains in FL accuracy and localization efficiency. Finally, future work will investigate the generalization of \tool beyond Java programs, extending to other widely used programming languages such as Python, C/C++, and JavaScript.

\section{Related Work}

\emph{Fault Localization} (FL) is essential to automated debugging, serving to identify program elements responsible for failures~\cite{yang2021evaluating,soremekun2021locating,wang_systematic_2025}. Effective FL reduces the search space for manual debugging and is used for automated program repair. 

\mypara{Spectrum-based Fault Localization (SBFL)} SBFL remains the baseline for automated debugging due to its efficiency and ease of deployment. Implementations such as the SFLKit workbench~\cite{smytzek2022sflkit} utilize execution traces to rank program elements based on their correlation with failure, while later approaches incorporate input data tracking to refine these rankings~\cite{demott_systematic_2013}. Correlation-based metrics such as Ochiai and Tarantula~\cite{jones2005empirical,abreu2006evaluation} struggle in the presence of multiple faults due to 
masking effects~\cite{abreu2007accuracy}, and their effectiveness is further reduced by coincidental correctness~\cite{masri2014prevalence}. Recent techniques attempt to isolate failure-inducing inputs~\cite{Kirschner:Soremekun:Zeller:2021}, but correlation-based approaches have a limited ability to distinguish real causes from mere correlations. 

\mypara{Mutation-based Fault Localization (MBFL)} MBFL refines SBFL by introducing active interventions, namely, injecting code changes (mutants) and observing their effect on the test outcomes. This provides stronger evidence of fault relevance than passive observation, and frameworks such as Major~\cite{major} have standardized mutant generation for Java programs. The evidential gains obtained by MBFL come at the cost of substantial computational overhead and the equivalent-mutant problem. 
 
Wang et al.~\cite{wang_systematic_2025} conduct a systematic investigation into different MBFL formulae and introduce a new method: BLMu. Although BLMu performs well on Defects4J, it requires a number of hyper-parameters, in particular two weighting coefficients on passing and failing tests respectively, $\alpha$ and $\beta$. 

\tool{} addresses a fundamental limitation of traditional MBFL approaches by replacing heuristic suspiciousness calculations with a formal notion of causal responsibility, weighted by semantic distance. Consequently, \tool{} provides not only evidence that a mutation is relevant to a failure, but an explanation of the degree to which that mutation is responsible for causing it.

\mypara{Causality in Software Testing} 
Baah et al.~\cite{baah_causal_2010} were the first to formulate fault localization as a causal inference problem, although their work primarily adjusted execution spectra. More recent approaches employ explicit counterfactual reasoning. CounterFault~\cite{podgurski2020counterfault} models faults by asking whether alternative variable values would have changed the outcome of execution. Further work integrated value- and predicate-based causal inference to isolate the causes of failure~\cite{kuccuk2021improving}.
While these approaches represent a significant step forward, they predominantly address \textit{data} causality (the state of program variables) rather than \textit{structural} causality (the program logic itself). 
Clark et al.~\cite{clark_testing_2023} argue for using causal inference to test scientific software.
In short, existing techniques either provide scalable statistical rankings without causal guarantees or causes in terms of
values of the variables (and not the program code).

\mypara{Semantic Distance and Program Repair} FL is often a precursor to \emph{Automated Program Repair} (APR), although the objectives differ. FL ranks suspicious elements, whereas APR seeks locations amenable to patch generation. Yang et al.~\cite{yang2021evaluating} show that highly ranked FL locations are not necessarily the most repairable, motivating metrics that account for repair cost and complexity. 
There is a substantial body of work on repair techniques based on semantic distance and
similarity~\cite{nguyen2013semfix,ke2015repairing,wang2005automated,feser2022metric,falleri2014fine,alon2019code2vec,eberlein2023semantic}.
Despite that, semantic distance has not been used for FL, and a potential connection of
causal responsibility to semantic significance of the corresponding code is unexplored.

\mypara{Large Language Models in Localization} LLMs have introduced new paradigms for FL~\cite{wu_large_2023}. 
Unfortunately, they remain susceptible to non-determinism and hallucination~\cite{chen_measure_2026}, hence reinforcing 
the need for deterministic, formally grounded FL techniques that provide interpretable and reproducible results. 

Statistical approaches provide scalable localization but lack causal guarantees; causal approaches reason about execution states rather than source-code responsibility; mutation-based approaches perform interventions but rely on heuristic ranking; and semantic-distance techniques quantify code similarity without contributing to causal attribution. Consequently, existing FL methods cannot formally determine which source-code changes are responsible for a failure, nor the degree of that responsibility. This gap motivates the development of causally grounded FL techniques that identify not only which code is a cause of a failure, but also how \emph{strongly} particular source-code changes contribute. \tool{} closes this gap by unifying intervention-based mutation analysis, actual causality, causal responsibility, and semantic distance within a single FL framework.

\section{Conclusions}
\label{sec:conc}
We introduced a new paradigm, CBFL, for fault localization, based on actual causality. This paradigm, CBFL, addresses the shortcomings of the existing statistical approaches by computing actual causes of faults and ranking them according to their degree of causal influence on the
fault. The problem of computing causes is intractable; hence we presented an algorithm and a number of heuristics for practical approximate computation of the ranking. We implemented this approach in a tool \tool, a causality-based fault-localization technique that uses responsibility to rank fault locations causally. We evaluated \tool on Defects4J using a common set of mutants and showed that it outperforms state-of-the-art MBFL techniques by 15\% in MRR and 16\% in Top-$1$. We further showed that guided mutation generation improves localization efficiency---in terms of mutants required to be evaluated---by $2.6\times$ over standard mutation strategies. These results show the potential of causal analysis for more accurate and actionable FL.




\newpage

\bibliographystyle{splncs04}
\bibliography{main.bib}
\newpage

\appendix
\section{Actual Causality}
The framework of actual causality~\cite{HP01b,halpern_actual_2016} describes the world in terms of variables and their values. A causal model $M$ consists of a set of variables $\V$ and a set of structural equations that capture dependencies between these variables. Some salient features of causal models can be represented by a \emph{causal network}: a directed acyclic graph whose nodes are the variables of $M$ and whose directed edges capture the existence and direction of causal dependence. 

A \emph{context} $\vec{u}$ is an assignment to the variables of $M$ that have no incoming edges in the causal network---that is, their values do not depend on the values of other variables. $\vec{u}$ determines, through the equations of $M$, the values of all variables in $\V$. 
Together, a causal model $M$ and a context $\vec{u}$, which specifies the particular situation being analyzed, form a causal setting $(M,\vec{u})$.
We write $(M,\vec{u}) \sat \varphi$ if a Boolean formula $\varphi$ holds in the causal setting $(M,\vec{u})$.

Causal analysis is performed through \emph{interventions}. An intervention $\vec{X}\leftarrow\vec{x}$ assigns values $\vec{x}$ to a subset $\vec{X}\subseteq\V$, replacing the corresponding equations with these assignments. We write $(M,\vec{u})\models[\vec{X}\leftarrow\vec{x}]\varphi$ if $\varphi$ holds after this intervention. Interventions, therefore, allow us to reason about the effect of changing selected variables on the outcome. With that, we can define actual causes:

\dfn[Actual cause~\cite{Hal15,halpern_actual_2016}]\label{def:AC}
$\vec{X} = \vec{x}$, a subset of $\V$ and their valuation, is 
an \emph{actual cause} of $\varphi$ in the causal setting $(M,\vec{u})$ if the following three conditions hold: 
\begin{description}
\item[{\rm AC1.}]\label{ac1} $(M,\vec{u}) \models (\vec{X} = \vec{x})$ and $(M,\vec{u}) \models \varphi $. 
\item[{\rm AC2.}]\label{ac2} There is an alternative setting $\vec{x}'$ of the variables in $\vec{X}$, a 
(possibly empty)  set $\vec{W}$ of variables in $\V - \vec{X}$, and a setting $\vec{w}$ of the variables in $\vec{W}$ such that $(M,\vec{u}) \models \vec{W} = \vec{w}$ and
$(M,\vec{u}) [\vec{X} \gets \vec{x}', \vec{W} \gets
    \vec{w}] \models \neg \varphi$.
\item[{\rm AC3.}] \label{ac3}\index{AC3}  
  $\vec{X}$ is minimal, i.e., there is no strict subset $\vec{X}'$ of
  $\vec{X}$ such that $\vec{X}' = \vec{x}''$ can replace $\vec{X} =
  \vec{x}'$ in 
  AC2, where $\vec{x}''$ is the restriction of
$\vec{x}'$ to the variables in $\vec{X}'$.
\end{description}
\edfn

Essentially, AC1 states that the cause, $\vec{X} = \vec{x}$, and the outcome, $\varphi$ indeed happened in context $\vec{u}$ (this is \emph{actual} causality). AC2 says that for $\vec{X} = \vec{x}$ to be a cause of $\varphi$, there must be an assignment $\vec{x}'$ for $\vec{X}$ such that if we clamp  $\vec{X} = \vec{x}'$ and  $\vec{W} =\vec{w}$,  $\varphi$ no longer holds. AC3 states that there should be no unnecessary variables in $\vec{X} = \vec{x}$. 
A variable $X$ in an actual cause $\vec{X}$ is called a \emph{part of a cause}. Following~\cite{Hal15}, we often refer to parts of causes as causes.

In this paper, we use the definition of \emph{degree of responsibility}, first introduced in~\cite{chockler2004responsibility}, which quantifies the measure of causal influence.

\dfn[Responsibility \cite{chockler2004responsibility}]\label{def:resp}
The \emph{degree of responsibility} of an endogenous variable $X$ for $\varphi$ in $(M,\vec{u})$ is defined as
$1/(|\vec{X}|+|\vec{W}|)$, where $\vec{X}$ is an actual cause for $\varphi$ containing $X$ that minimizes $|\vec{X}|+|\vec{W}|$, and $\vec{W}$ is its respective contingency.
If there is no actual cause containing $X$, its degree of responsibility is $0$.
\edfn

\section{Experimental Details}
\Cref{tab:structuraldist} details the heuristic distance weights applied during Experiment 2 to approximate structural mutation distances based on AST node types.
\begin{table}[htb!]
    \centering
    \caption{Heuristic structural distance function.}
    \footnotesize
    \begin{tabular}{l|l|c}
    \toprule
        \textbf{Mutation Type of $m$} & \textbf{Target Element $s$ Type} & Structural Distance \\
        \midrule
         Deletion & Function  &  1.0 \\
         Deletion & Assignment / Control Flow / Statements  &  0.8 \\
         Replacement & Conditional & 0.6 \\
         Replacement & Arithmetic / Relational & 0.4 \\
         Modification & Literal Value / Variable & 0.2 \\
    \end{tabular}
    \label{tab:structuraldist}
\end{table}
\section{Detailed Results}
\Cref{tab:average_metrics} summarizes the experimental metadata across the 17 projects, detailing the average number of mutants generated, test suite sizes, and the number of failing and passing test within the test suite for both experiments. This 

\begin{table}[h!]
\centering
\caption{Average test suite execution and mutant generation statistics across evaluated Defects4J projects for Experiment 1 (exhaustive) and Experiment 2 (heuristic search). This is with respect to the versions ran for each experiment.}
\label{tab:average_metrics}
\begin{tabular}{llrrrr}
\toprule
\textbf{Exp.} & \textbf{Project} & \textbf{Mutants} & \textbf{Total Tests} & \textbf{Failed} & \textbf{Passed} \\
\midrule
\multirow[t]{17}{*}{1} 
& Chart & 570.25 & 1819.79 & 3.75 & 1816.04 \\
 & Cli & 175.28 & 263.56 & 1.69 & 261.87 \\
 & Closure & 520.19 & 6910.36 & 3.24 & 6907.12 \\
 & Codec & 707.83 & 447.06 & 2.39 & 444.67 \\
 & Collections & 222.46 & 13149.89 & 1.36 & 13148.54 \\
 & Compress & 490.43 & 434.43 & 1.53 & 432.89 \\
 & Csv & 230.50 & 197.81 & 1.50 & 196.31 \\
 & Gson & 1249.00 & 986.20 & 1.93 & 984.27 \\
 & JacksonCore & 1904.23 & 356.19 & 2.04 & 354.15 \\
 & JacksonDatabind & 240.05 & 1619.64 & 1.18 & 1618.45 \\
 & JacksonXml & 237.17 & 152.17 & 2.00 & 150.17 \\
 & Jsoup & 563.15 & 453.60 & 1.56 & 452.04 \\
 & JxPath & 286.45 & 346.68 & 1.68 & 345.00 \\
 & Lang & 810.16 & 1814.18 & 1.97 & 1812.21 \\
 & Math & 847.51 & 2508.55 & 1.67 & 2506.88 \\
 & Mockito & 32.74 & 1137.31 & 2.97 & 1134.34 \\
 & Time & 508.36 & 3916.84 & 2.64 & 3914.20 \\
\midrule
\multirow[t]{5}{*}{2} 
    & Csv & \textbf{87.09} & 193.55 & 1.55 & 192.00 \\
 & Gson & 95.50 & 852.00 & 1.00 & 851.00 \\
 & JacksonXml & 167.17 & \textbf{152.17} & 2.00 & 150.17 \\
 & Math & 134.45 & 2505.30 & 1.53 & 2503.77 \\
\bottomrule
\end{tabular}
\end{table}

\Cref{tab:global_metrics_exp1_2} and \Cref{tab:exp1_ablation} present the complete global fault localization metrics and ablation study data. Furthermore, \Cref{tab:exp1_ablation} (Sensitivity Analysis) demonstrates the robustness of the $\lambda$ parameter, confirming that \tool{}'s semantic distance weighting remains stable across a wide range of values without overfitting.
\begin{table}[hbt!]
    \centering
    \caption{Global fault localization effectiveness across both experimental setups.}
    \label{tab:global_metrics_exp1_2}
    \footnotesize
    \begin{tabular}{clrrrrrr}
        \toprule
        \textbf{Exp.} & \textbf{Method} & \textbf{Top-1} & \textbf{Top-3} & \textbf{Top-5} & \textbf{Top-10} & \textbf{MAR} & \textbf{MRR} \\
        \midrule
        \multirow{8}{*}{Major (1)}
&\tool{} & \textbf{103} & \textbf{203} & \textbf{244} & \textbf{301} & 48.132 & \textbf{0.349} \\
&\texttt{BLMu} & 89 & 168 & 211 & 270 & \textbf{45.578} & 0.304 \\
&\texttt{Barinel} &  71 & 122 & 140 & 176 & 81.575 & 0.222 \\
&\texttt{Metallaxis} & 40 & 93 & 122 & 185 & 61.652 & 0.181 \\
&\texttt{MUSE} &  68 & 120 & 139 & 187 & 68.609 & 0.224 \\
&\texttt{Tarantula} &67 & 113 & 132 & 171 & 81.705 & 0.213 \\
&\texttt{DStar2} & 63 & 123 & 140 & 176 & 81.571 & 0.218 \\
&\texttt{jaccard} & 59 & 113 & 131 & 170 & 81.751 & 0.204 \\
&\texttt{Op2} & 23 & 44 & 56 & 117 & 76.218 & 0.113 \\
\midrule
\multirow{8}{*}{\tool{} (2)}
&\tool{} & \textbf{10} & \textbf{19} & \textbf{23} & \textbf{25} & 135.648 & \textbf{0.223}\\
& \texttt{BLMu} & 7 & 14 & 19 & 24 & \textbf{88.840} & 0.163 \\
&\texttt{Barinel} & 5 & 9 & 10 & 11 & 99.488 & 0.103 \\
&\texttt{Metallaxis} & 2 & 9 & 14 & 21 & 92.660 & 0.112 \\
&\texttt{Muse} & 5 & 9 & 10 & 11 & 98.025 & 0.103 \\
&\texttt{Tarantula} & 6 & 9 & 10 & 11 & 99.494 & 0.108 \\
&\texttt{DStar2} & 3 & 6 & 10 & 12 & 99.500 & 0.083 \\
&\texttt{Jaccard}  & 6 & 9 & 10 & 11 & 99.506 & 0.107 \\
&\texttt{Op2} & 5 & 8 & 9 & 10 & 99.228 & 0.097 \\
\bottomrule
    \end{tabular}
\end{table}

\begin{table}[hbt!]
    \centering
    \caption{Ablation Study - Experiment 1.}
    \footnotesize
    \begin{tabular}{lrrrrrrr}
        \toprule
        \textbf{Configuration} & \textbf{Top-1} & \textbf{Top-3} & \textbf{Top-5} & \textbf{Top-10} & \textbf{MAR} & \textbf{MRR} \\
        \midrule
         \rexbase{} & 50 & 164 & 215 & 284 & 45.903 & 0.269 \\
        \midrule
        \rexpenalty{} & 75 & 195 & \textbf{259} & \textbf{311} & 47.605 & 0.335 \\
        \rexdist{} (Heuristic) & 65 & 174 & 226 & 289 & \textbf{45.260} & 0.297 \\
        \rexdistsem{} (CodeBERT) &93 & 193 & 237 & 292 & 48.784 & 0.331 \\
         \midrule
        \rexpendist{} & 81 & 196 & 248 & 309 & 48.306 & 0.331 \\
        \rexpendistsem{} & \textbf{103} & \textbf{203} & 244 & 301 & 48.132 & \textbf{0.349} \\
        \bottomrule
    \end{tabular}
    \label{tab:exp1_ablation}
\end{table}

\begin{table}[hbt!]
    \centering
    \caption{Ablation Study - Experiment 2.}
    \footnotesize
    \begin{tabular}{lrrrrrrr}
        \toprule
        \textbf{Configuration} & \textbf{Top-1} & \textbf{Top-3} & \textbf{Top-5} & \textbf{Top-10} & \textbf{MAR} & \textbf{MRR} \\
        \midrule
         \rexbase{} & 3 & 5 & 7 & 11 & 207.623 & 0.084 \\
        \midrule
        \rexpenalty{}& 6 & 13 & 17 & 21 & \textbf{133.457} & 0.160 \\
        \rexdist{} (Heuristic) & 3 & 5 & 7 & 11 & 207.623 & 0.084 \\
        \rexdistsem{} (CodeBERT) & \textbf{10} & 17 & 21 & 27 & 198.574 & 0.208 \\
         \midrule
        \rexpendist{} & 6 & 13 & 17 & 21 & \textbf{133.457} & 0.160 \\
        \rexpendistsem{} & \textbf{10} & \textbf{20} & \textbf{24} & \textbf{26} & 135.648 & \textbf{0.223} \\
        \bottomrule
    \end{tabular}
    \label{tab:exp1_ablation}
\end{table}


\begin{table}[hbt!]
    \centering
     \caption{Sensitivity Analysis: $\lambda$ for Experiment 1}
    \resizebox{0.9\textwidth}{!}{
    \centering
    \begin{subfigure}[t]{0.45\textwidth}
    \begin{tabular}{lrrrrrrr}
        \toprule
        \textbf{$\lambda$} & \textbf{Top-1} & \textbf{Top-3} & \textbf{Top-5} & \textbf{Top-10} & \textbf{MAR} & \textbf{MRR} \\
        \midrule
         0.05 & 103 & 203 & 244 & 301 & 48.132 & 0.349 \\
        0.1 & 102 & 204 & 247 & 299 & 48.171 & 0.347 \\
        0.25 & 100 & 202 & 245 & 297 & 48.291 & 0.343 \\
        0.5 & 99 & 199 & 242 & 297 & 48.403 & 0.341 \\
        1 & 98 & 196 & 243 & 296 & 48.566 & 0.339 \\
        2 & 98 & 196 & 241 & 293 & 48.676 & 0.339 \\
        3 & 97 & 196 & 241 & 294 & 48.694 & 0.338 \\
        \bottomrule
    \end{tabular}
    \caption{\rexpendistsem{}}
    \end{subfigure}
    \hspace{1.6cm}
    \centering
    \begin{subfigure}[t]{0.45\textwidth}
    \begin{tabular}{lrrrrrr}
        \toprule
        \textbf{$\lambda$} & \textbf{Top-1} & \textbf{Top-3} & \textbf{Top-5} & \textbf{Top-10} & \textbf{MAR} & \textbf{MRR} \\
        \midrule
         0.05  & 81 & 196 & 248 & 309 & 48.306 & 0.331 \\
        0.1 & 81 & 196 & 248 & 309 & 48.308 & 0.330 \\
        0.25 & 81 & 196 & 248 & 309 & 48.308 & 0.330 \\
        0.5 & 81 & 196 & 248 & 309 & 48.308 & 0.330 \\
        1 & 81 & 196 & 248 & 309 & 48.308 & 0.330 \\
        2 & 78 & 198 & 246 & 307 & 48.476 & 0.327 \\
        3 & 75 & 196 & 244 & 306 & 48.899 & 0.321 \\
        \bottomrule
    \end{tabular}
    \caption{\rexpendist{}}
    \end{subfigure}}
    \label{tab:exp1_ablation}
\end{table}

To ensure complete transparency regarding project-level variance, we provide absolute Top-$N$ accuracy counts ($N \in \{1, 3, 5, 10\}$) for all baseline methods across every evaluated project (\Cref{tab:project_top1}--\Cref{tab:project_top10}).

\begin{table}[htb!]
    \centering
    \caption{Project-level comparison of Top-1 accuracy (absolute number of bugs located).}
    \label{tab:project_top1}
    \resizebox{\columnwidth}{!}{%
\begin{tabular}{lrrrrrrrrr}
\toprule
Project & SoftReX & Barinel & MUSE & DStar2 & Tarantula & Metallaxis & Jaccard & Op2 & BLMu \\
\midrule
Chart & 5 & 2 & 2 & 3 & 2 & 0 & 2 & 1 & 2 \\
Cli & 3 & 4 & 3 & 1 & 3 & 6 & 2 & 4 & 5 \\
Closure & 14 & 0 & 0 & 0 & 0 & 0 & 0 & 0 & 0 \\
Codec & 2 & 2 & 2 & 1 & 3 & 0 & 1 & 1 & 2 \\
Collections & 2 & 1 & 2 & 2 & 1 & 2 & 2 & 1 & 3 \\
Compress & 6 & 5 & 5 & 5 & 5 & 3 & 5 & 2 & 6 \\
Csv & 4 & 1 & 1 & 1 & 1 & 0 & 1 & 0 & 1 \\
Gson & 1 & 1 & 1 & 1 & 1 & 1 & 1 & 0 & 3 \\
JacksonCore & 2 & 1 & 1 & 1 & 1 & 0 & 1 & 0 & 1 \\
JacksonDatabind & 12 & 8 & 9 & 8 & 9 & 6 & 9 & 2 & 14 \\
JacksonXml & 0 & 0 & 0 & 0 & 0 & 1 & 0 & 0 & 1 \\
Jsoup & 13 & 11 & 11 & 7 & 11 & 8 & 9 & 4 & 15 \\
JxPath & 6 & 3 & 1 & 4 & 1 & 1 & 1 & 1 & 1 \\
Lang & 7 & 11 & 11 & 11 & 11 & 2 & 8 & 1 & 11 \\
Math & 17 & 15 & 13 & 12 & 12 & 4 & 12 & 4 & 16 \\
Mockito & 4 & 3 & 3 & 3 & 3 & 4 & 2 & 0 & 4 \\
Time & 5 & 3 & 3 & 3 & 3 & 2 & 3 & 2 & 4 \\
\bottomrule
\end{tabular}

    }
\end{table}
\begin{table}[htb!]
    \centering
    \caption{Project-level comparison of Top-3 accuracy (absolute number of bugs located).}
    \label{tab:project_top3}
    \resizebox{\columnwidth}{!}{%
\begin{tabular}{lrrrrrrrrr}
\toprule
Project & SoftReX & Barinel & MUSE & DStar2 & Tarantula & Metallaxis & Jaccard & Op2  & BLMu \\
\midrule
Chart & 5 & 4 & 5 & 4 & 4 & 1 & 4 & 4 & 4 \\
Cli & 11 & 6 & 7 & 7 & 5 & 11 & 5 & 4 & 11 \\
Closure & 37 & 0 & 0 & 0 & 0 & 0 & 0 & 1 & 0 \\
Codec & 4 & 4 & 5 & 3 & 4 & 0 & 4 & 3 & 4 \\
Collections & 4 & 3 & 3 & 3 & 3 & 2 & 3 & 1 & 5 \\
Compress & 14 & 11 & 10 & 12 & 10 & 6 & 10 & 3 & 14 \\
Csv & 5 & 4 & 4 & 4 & 3 & 2 & 3 & 1 & 4 \\
Gson & 3 & 1 & 1 & 1 & 1 & 4 & 1 & 1 & 4 \\
JacksonCore & 4 & 4 & 4 & 5 & 3 & 4 & 4 & 1 & 7 \\
JacksonDatabind & 23 & 16 & 15 & 15 & 15 & 12 & 16 & 4 & 24 \\
JacksonXml & 0 & 0 & 0 & 0 & 0 & 2 & 0 & 0 & 2 \\
Jsoup & 20 & 16 & 15 & 15 & 15 & 15 & 13 & 6 & 21 \\
JxPath & 10 & 6 & 5 & 6 & 5 & 4 & 5 & 1 & 8 \\
Lang & 18 & 14 & 14 & 15 & 14 & 11 & 13 & 2 & 19 \\
Math & 28 & 23 & 22 & 23 & 21 & 10 & 22 & 8 & 26 \\
Mockito & 8 & 5 & 5 & 5 & 5 & 6 & 5 & 2 & 9 \\
Time & 9 & 5 & 5 & 5 & 5 & 3 & 5 & 2 & 6 \\
\bottomrule
\end{tabular}

    }
\end{table}
\begin{table}[htb!]
    \centering
    \caption{Project-level comparison of Top-5 accuracy (absolute number of bugs located).}
    \label{tab:project_top5}
    \resizebox{\columnwidth}{!}{%
\begin{tabular}{lrrrrrrrrr}
\toprule
Project & SoftReX & Barinel & MUSE & DStar2 & Tarantula & Metallaxis & Jaccard & Op2 & BLMu \\
\midrule
Chart & 6 & 4 & 5 & 4 & 4 & 1 & 4 & 4 & 6 \\
Cli & 12 & 7 & 7 & 8 & 7 & 13 & 7 & 4 & 13 \\
Closure & 49 & 0 & 1 & 0 & 0 & 0 & 0 & 3 & 0 \\
Codec & 5 & 5 & 6 & 4 & 4 & 2 & 4 & 3 & 5 \\
Collections & 5 & 3 & 3 & 3 & 3 & 2 & 3 & 1 & 5 \\
Compress & 16 & 13 & 13 & 13 & 13 & 6 & 13 & 4 & 17 \\
Csv & 7 & 4 & 4 & 4 & 3 & 2 & 3 & 2 & 6 \\
Gson & 5 & 1 & 1 & 1 & 1 & 4 & 1 & 1 & 4 \\
JacksonCore & 5 & 5 & 4 & 5 & 4 & 5 & 4 & 1 & 8 \\
JacksonDatabind & 27 & 18 & 18 & 18 & 17 & 14 & 17 & 6 & 28 \\
JacksonXml & 2 & 0 & 0 & 0 & 0 & 3 & 0 & 0 & 3 \\
Jsoup & 21 & 18 & 18 & 19 & 17 & 21 & 16 & 9 & 29 \\
JxPath & 10 & 7 & 7 & 7 & 7 & 5 & 7 & 1 & 10 \\
Lang & 20 & 15 & 15 & 15 & 15 & 17 & 14 & 2 & 24 \\
Math & 32 & 27 & 26 & 26 & 24 & 13 & 25 & 10 & 34 \\
Mockito & 12 & 8 & 6 & 8 & 8 & 10 & 8 & 3 & 12 \\
Time & 10 & 5 & 5 & 5 & 5 & 4 & 5 & 2 & 7 \\
\bottomrule
\end{tabular}

    }
\end{table}
\begin{table}[htb!]
    \centering
    \caption{Project-level comparison of Top-10 accuracy (absolute number of bugs located).}
    \label{tab:project_top10}
    \small
\begin{tabular}{lrrrrrrrrr}
\toprule
Project & SoftReX & Barinel & MUSE & DStar2 & Tarantula & Metallaxis & Jaccard & Op2 & BLMu \\
\midrule
Chart & 7 & 6 & 5 & 6 & 5 & 4 & 5 & 4 & 8 \\
Cli & 13 & 11 & 11 & 11 & 10 & 15 & 10 & 9 & 17 \\
Closure & 55 & 0 & 5 & 0 & 0 & 1 & 0 & 15 & 2 \\
Codec & 8 & 6 & 7 & 6 & 6 & 3 & 6 & 6 & 7 \\
Collections & 5 & 3 & 3 & 3 & 3 & 2 & 3 & 1 & 6 \\
Compress & 20 & 14 & 16 & 14 & 14 & 11 & 14 & 7 & 19 \\
Csv & 8 & 5 & 5 & 5 & 5 & 5 & 5 & 2 & 8 \\
Gson & 5 & 3 & 2 & 3 & 3 & 6 & 3 & 2 & 5 \\
JacksonCore & 7 & 5 & 5 & 5 & 5 & 8 & 5 & 1 & 10 \\
JacksonDatabind & 41 & 25 & 21 & 25 & 25 & 26 & 25 & 12 & 37 \\
JacksonXml & 3 & 0 & 0 & 0 & 0 & 3 & 0 & 0 & 3 \\
Jsoup & 28 & 20 & 28 & 20 & 19 & 26 & 19 & 17 & 37 \\
JxPath & 11 & 8 & 7 & 8 & 8 & 7 & 8 & 1 & 11 \\
Lang & 29 & 20 & 19 & 20 & 19 & 23 & 18 & 8 & 31 \\
Math & 34 & 31 & 35 & 31 & 30 & 24 & 30 & 18 & 44 \\
Mockito & 15 & 14 & 13 & 14 & 14 & 14 & 14 & 12 & 14 \\
Time & 12 & 5 & 5 & 5 & 5 & 7 & 5 & 2 & 11 \\
\bottomrule
\end{tabular}
\end{table}

\end{document}